%% file: main.tex
\documentclass[a4paper,10pt,fleqn,usenames,dvipsnames]{article}

\input{preamble}

\input{includes/definitions}

\usepackage[nonatbib,preprint]{neurips}

\title{Accelerating template generation in resonant anomaly detection searches with optimal transport}
\author{%
\quad\quad\quad Matthew Leigh\\
\quad\quad\quad University of Geneva\\
\quad\quad\quad \texttt{matthew.leigh@unige.ch}\\
\And
\quad\quad\quad Debajyoti Sengupta\\
\quad\quad\quad University of Geneva\\
\quad\quad\quad \texttt{debajyoti.sengupta@unige.ch}\\
\And
Benjamin Nachman\\
Lawrence Berkeley National Laboratory\\
\texttt{bpnachman@lbl.gov}
\And
Tobias Golling\\
University of Geneva\\
\texttt{tobias.golling@unige.ch}
}

\begin{document}
\maketitle

\begin{abstract}
    We introduce Resonant Anomaly Detection with Optimal Transport (\oli), a method for generating signal templates in resonant anomaly detection searches.
    \oli leverages the fact that the conditional probability density of the target features vary approximately linearly along the optimal transport path connecting the resonant feature.  This does not assume that the conditional density itself is linear with the resonant feature, allowing \oli to efficiently capture multimodal relationships, changes in resolution, etc.
    By solving the optimal transport problem, \oli can quickly build a template by interpolating between the background distributions in two sideband regions.
    We demonstrate the performance of \oli using the LHC Olympics R\&D dataset, where we find comparable sensitivity and improved stability with respect to deep learning-based approaches.
\end{abstract}

\section{Introduction}

\input{tex/intro}

\section{Dataset}
\label{sec:dataset}
\input{tex/data}
\section{Template Building from Optimal Interpolants}
\label{sec:methods}
\input{tex/method}

\section{Anomaly Detection on the LHCO}
\label{sec:results}
\input{tex/results}

\section{Discussion}
\label{sec:conclusions}
\input{tex/discussion}

\section*{Acknowledgements}
\input{includes/acknowledgement}

\phantomsection
\addcontentsline{toc}{chapter}{References}
\printbibliography[title=References]

\end{document}

%% file: preamble.tex
\usepackage{lmodern}
\usepackage[T1]{fontenc}
\usepackage{amsmath}
\usepackage{amssymb}

\usepackage{url}
\usepackage{longtable}
\usepackage{multirow}
\usepackage{caption}
\usepackage[labelformat=simple]{subcaption}

\usepackage[nodayofweek]{datetime}
\usepackage{enumerate}
\usepackage{xspace}
\usepackage{xcolor}
\usepackage{graphicx}
\usepackage[pdftex,hidelinks]{hyperref}
\usepackage{bookmark}
\usepackage{booktabs}
\usepackage{algpseudocode}
\usepackage{algorithm}
\usepackage{appendix}

\usepackage[normalem]{ulem}
\usepackage{tabularx}

\newcolumntype{Y}{>{\centering\arraybackslash}X}
\newcolumntype{R}{>{\raggedleft\arraybackslash}X}

\usepackage{float}
\usepackage[capitalise]{cleveref}
\usepackage{lipsum}
\usepackage[backend=biber,giveninits=true,doi=false,sorting=none,style=numeric-comp]{biblatex}
\DeclareFieldFormat[article]{citetitle}{\textit{#1}\isdot}
\DeclareFieldFormat[article]{title}{\textit{#1}\isdot}
\DeclareFieldFormat[unpublished]{citetitle}{\textit{#1}\isdot}
\DeclareFieldFormat[unpublished]{title}{\textit{#1}\isdot}
\DeclareFieldFormat[inproceedings]{citetitle}{\textit{#1}\isdot}
\DeclareFieldFormat[inproceedings]{title}{\textit{#1}\isdot}

\DeclareNameAlias{author}{last-first}

%% file: includes/definitions.tex
\newcommand{\SR}{SR\xspace}
\newcommand{\SBone}{SB1\xspace}
\newcommand{\SBtwo}{SB2\xspace}

\newcommand{\FfF}{\mbox{\textsc{Curtain}sF4F}\xspace}

\newcommand{\cwola}{\mbox{\textsc{CWoLa}}\xspace}
\newcommand{\mjj}{\ensuremath{m_{JJ}}\xspace}

\newcommand{\oli}{\textsc{RAD-OT}\xspace}

%% file: tex/intro.tex
The Standard Model (SM) has been very successful in describing the fundamental particles and their interactions, but there are many reasons why it is not the final theory of nature, such as the unexplained dark matter in the universe.
One of the main goals of the Large Hadron Collider (LHC) is to search for new physics beyond the Standard Model (BSM) of particle physics.
General purpose detectors such as ATLAS~\cite{ATLAS:2008xda} and CMS~\cite{CMS:2008xjf} are designed to be sensitive to a wide range of new physics possibilities.
As there are limitless possibilities as to what the new physics might be, it is not feasible to individually test each hypothesis.
It is thus desirable to have methods that are simultaneously sensitive to numerous possibilities~\cite{Kasieczka:2021xcg,Aarrestad:2021oeb,Karagiorgi:2021ngt}.  This would complement and could be combined with dedicated searches in specific regions of phase space.
A number of data-driven methods~\cite{Collins:2018epr,cwolabump,anode,Amram_2021,Stein:2020rou,Kamenik:2022qxs,Andreassen:2020nkr,Benkendorfer:2020gek,cathode,curtains,curtainsf4f,feta,Chen:2022suv,drapes,ranode,Golling:2023yjq,Metodiev:2023izu,Buhmann:2023acn} have been developed and applied~\cite{ATLAS:2020iwa,Shih:2021kbt,Shih:2023jfv,Pettee:2023zra,CMS-PAS-EXO-22-026,Sengupta:2024ezl} in the context of resonant anomaly searches\footnote{We are not counting generic anomaly detection methods applied to the resonant case, see e.g. the recent ATLAS results~\cite{ATLAS:2023azi,ATLAS-CONF-2023-022} and method papers they cite.}.
The main assumption is that the new physics is localised in a known (resonant) feature, while the background distribution is featureless.  Data away from the resonance in the sideband (SB) are used to build an unbinned template of the background distribution under the resonance in the signal region via interpolation.  The most widely-studied approaches use conditional discriminative or generative machine learning models. Once the template is built, it is compared to the data in the SR, often using a classifier to create an anomaly score~\cite{cwola,Collins:2018epr,cwolabump}.

Existing methods have shown promising results, but there are a number of motivations for building new techniques.  For example, neural network-based methods can be computationally expensive.  This is especially the case when ensembling is required for stability to minimize fluctuations from the scholastic nature of training. In this work, we propose a template generation strategy that does not rely on neural networks in order to accelerate the process and enhance stability while also preserving sensitivity.  The strategy uses the framework of Optimal Transport (OT), a set of tools for transforming one dataset into another with the least amount of movement.  OT has been studied as a method for creating an anomaly score~\cite{CrispimRomao:2020ejk,Fraser:2021lxm,Park:2022zov,Craig:2024rlv}, but we propose to use it for template generation.   Our Resonant Anomaly Detection with Optimal Transport (\oli) leverages the fact that the conditional probability density of the target features vary approximately linearly along the OT path connecting the resonant feature.  This does not assume that the conditional density itself is linear with the resonant feature, allowing \oli to efficiently capture multimodal relationships, changes in resolution, etc.  For modest feature space dimensionality, the OT solution can be approximated without neural networks\footnote{See Ref.~\cite{curtains} for OT-related approaches based on neural networks.}, leading to high efficiency  and stability.

This paper is organized as follows.  Section~\ref{sec:dataset} briefly reviews the LHC Olympics (LHCO) R\&D dataset~\cite{LHCOlympics} that we use to demonstrate \oli.  The mathematical aspects of \oli are described in Sec.~\ref{sec:methods}.  Numerical results on the LHC are presented in Sec.~\ref{sec:results}, and the paper ends with conclusions and outlook in Sec.~\ref{sec:conclusions}.

%% file: tex/data.tex
The LHCO R\&D dataset~\cite{LHCOlympics} comprises background events represented by quark/gluon scattering to produce dijets and signal events arising from the all-hadronic decay of a massive particle to two other massive particles
$W^\prime\rightarrow X(\rightarrow q\bar{q}) Y(\rightarrow q\bar{q})$, 
with masses $m_{W^\prime} = 3.5$~TeV, $m_{X} = 500$~GeV, and $m_{Y} = 100$~GeV.
Both processes are simulated with \texttt{Pythia}~8.219~\cite{Sjostrand:2014zea} and interfaced to \texttt{Delphes}~3.4.1~\cite{deFavereau:2013fsa} for detector simulation.
Jets are reconstructed using the anti-$k_\mathrm{T}$ clustering algorithm~\cite{AntiKt} with a radius parameter $R=1.0$, using the \texttt{FastJet}~\cite{Cacciari:2011ma} package.
In total there are 1~million dijet events and $100,000$ signal events.

Events are required to have at least one $R = 1.0$ jet $J$ with pseudrapidity $\left| \eta \right| < 2.5 $, and transverse momentum $p_\mathrm{T}^{J} > 1.2$~TeV.
The top two leading $p_\mathrm{T}$ jets are selected and ordered by decreasing mass; they are labelled $J_1$ and $J_2$.
In order to remove the turn on in the \mjj distribution arising from the jet selections, we only consider events with $\mjj > 2.8$~TeV.
To construct the training datasets, we use varying amounts of signal events mixed in with the dijet events.

To study the performance of \oli, we use the same high-level features employed by many of the existing methods demonstrated on the same dataset.
These features are $m_{JJ}, m_{J_1}, \Delta m_{J} = m_{J_1} - m_{J_2}, \tau_{21}^{J_1}, \tau_{21}^{J_2}, \mathrm{and} \, \Delta R_{JJ}$, where $\tau_{21}^J$ is the N-subjettinness~\cite{nsubjettiness} ratio of jet $J$,
and $\Delta R_{JJ}$ is the angular separation of the two jets in the detector $\eta-\phi$ space.

%% file: tex/method.tex
\begin{figure}[hbpt]
    \centering
    \scalebox{0.6}{
        \includegraphics[width=1.\textwidth]{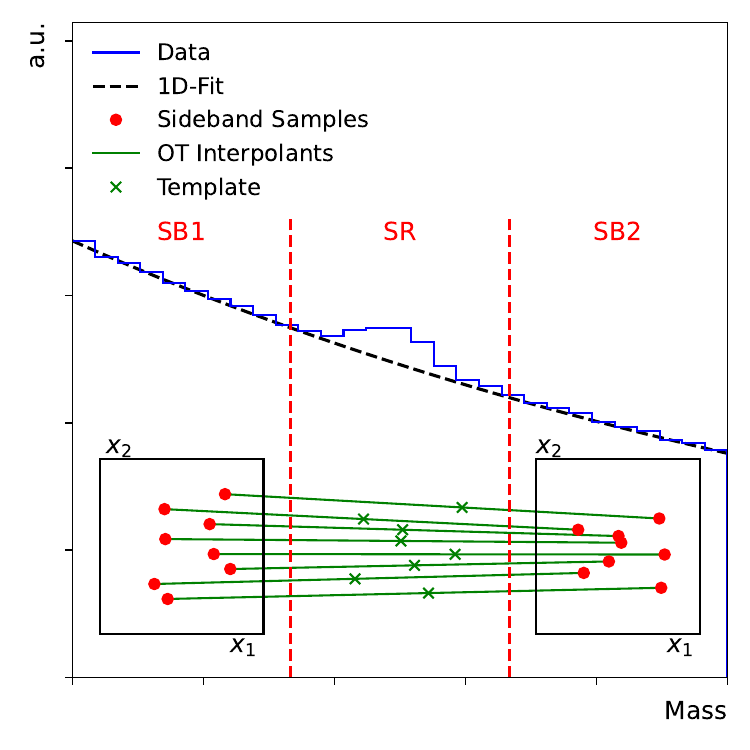}
    }
    \caption{A simple diagrammatic representation of the \oli method.
        Features $\mathbf{x}$ (red) are sampled from the mass sidebands \SBone and \SBtwo and paired using the optimal transport map (green).
        The resonant feature $m$ is sampled from the Kernel Density Estimator (KDE) (dashed black) and the SR template (green cross) is generated using linear interpolation.}
    \label{fig:method}
\end{figure}

This section describes the procedure for building a data-driven template using \oli.
A diagrammatic representation of the method is shown in~\autoref{fig:method}.

First, the features $\mathbf{f}$ of a dataset are split into a selected resonant feature $m$ and additional attributes that characterise the template, $\mathbf{x}$.
The data is then partitioned on $m$ into a signal region (SR) and two flanking sideband regions, \SBone and \SBtwo.
The goal is to sample from values of $\mathbf{f} = (m, \mathbf{x})$ that are representative of the background distribution in the SR.
Here, the assumption is that the population of the sidebands is predominantly background.
Furthermore, this method assumes that the conditional probability density of the additional features vary linearly along the optimal transport path connecting $m$. While this assumption may not hold exactly, it is a reasonable first order approximation to apply to narrow SRs.  For our numerical tests, we define our \SR based on a mass cut of $3300\leq\mjj<3700$~GeV, and sideband regions $3100\leq\mjj<3300$~GeV and $3700\leq\mjj<3900$~GeV.  In practice, these regions would be swept across the spectrum, but this region was chosen, as in previous papers, as it is centered near the LHCO signal mass peak.

Next, a one-dimensional fit is performed on the $m$ distribution to approximate $p(m)$~\cite{cathode}.
This can be done with a parametric or non-parametric (such as a Kernel Density Estimator, KDE) fit with just the sidebands or on all data.
This fit is then used to sample the mass values in the SR.

The next step is to build a linear interpolation paths between samples drawn from each sideband.
For $\mathbf{f}_1 = (m_1, \mathbf{x}_1) \in$ \SBone and $\mathbf{f}_2 = (m_2, \mathbf{x}_2) \in$ \SBtwo, this sets up the basic parametric function
\begin{equation}
    \label{eq:linear_interp}
    \begin{aligned}
        \mathbf{f}_t = \left( m_t, \mathbf{x}_1 + \frac{m_t - m_1}{m_2 - m_1} (\mathbf{x}_2 - \mathbf{x}_1) \right), \\
    \end{aligned}
\end{equation}
where $m_t$ is the desired mass value in the SR.

\begin{figure}[hbpt]
    \centering
    \includegraphics[width=0.48\textwidth]{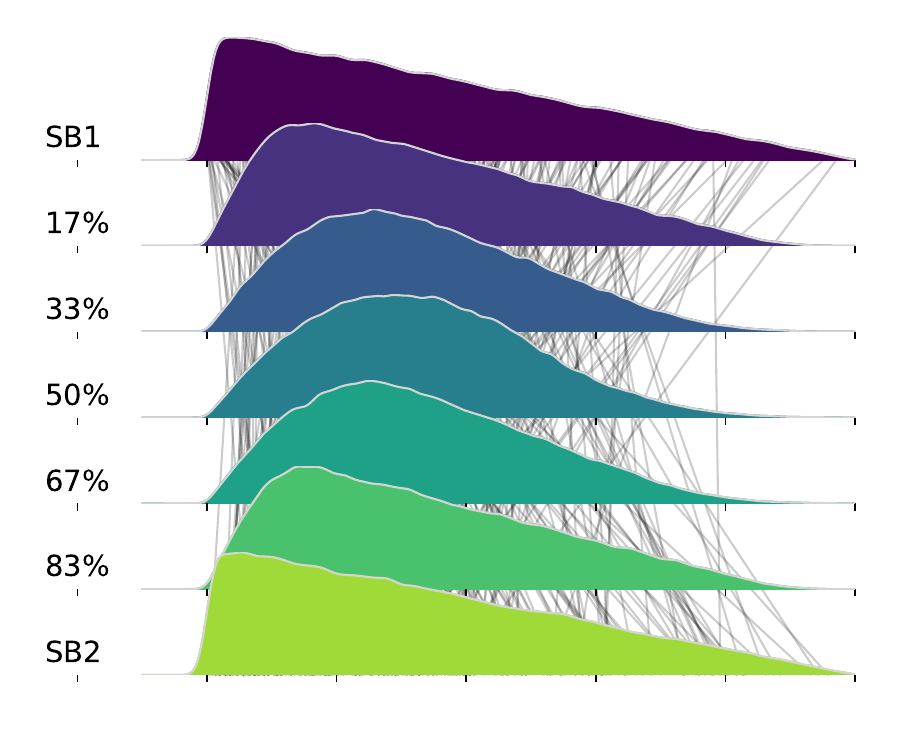}
    \includegraphics[width=0.48\textwidth]{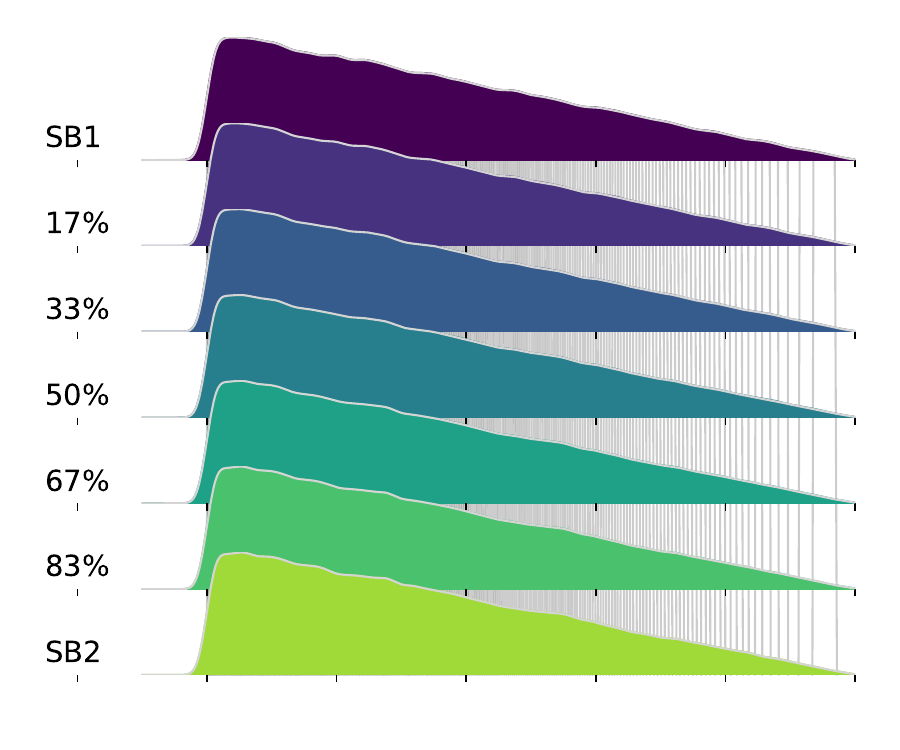}
    \caption{An illustration of the interpolated distributions with and without optimal transport matching on a one-dimensional toy dataset.
    The y-axis shows the slices of the interpolation between SB1~(0\%) and SB2~(100\%). Several interpolation paths are shown as black lines. In this toy sample the underlying distributions in each sideband is the same, and therefore the ideal interpolations should be constant. Without OT-matching (left), the interpolations result in a basic convolution and the intermediate stages no longer match the endpoints. The OT-matched interpolation paths (right) correctly keep the distribution constant from sideband to sideband.}
    \label{fig:toy}
\end{figure}

The crucial aspect of this method is to pair the samples $f_1$ and $f_2$ such that the interpolation is meaningful.
For instance, if the samples are drawn randomly and independently, the interpolated features will be pushed towards a normal distribution as illustrated in \cref{fig:toy}.

We propose to select pairs of samples $(f_1, f_2)$ such that they solve an optimal transport problem.
More specifically, given $N$ samples drawn from each sideband, we can construct a cost matrix $C$ where $C_{ij} = c(\mathbf{x}_1^i, \mathbf{x}_2^j)$ is the cost of transporting $\mathbf{x}_1^i$ to $\mathbf{x}_2^j$, where $i,j = 1, \ldots, N$.
We can then learn the map $\Gamma^*$ that minimises the cost of transporting the samples from \SBone to \SBtwo, by solving the following optimization problem:

\begin{equation}
    \centering
    \label{eq:optimal_transport}
    \begin{aligned}
        \Gamma^* = \min_{\Gamma} & \sum_{i=1}^N \sum_{j=1}^N \Gamma_{ij} c(\mathbf{x}_1^i, \mathbf{x}_2^j)       \\
        \text{subject to}        & \sum_{j=1}^N \Gamma_{ij} = p_i, \quad \forall i = 1, \ldots, n,               \\
                                 & \sum_{i=1}^N \Gamma_{ij} = q_j, \quad \forall j = 1, \ldots, m,               \\
                                 & \Gamma_{ij} \geq 0, \quad \forall i = 1, \ldots, N, \forall j = 1, \ldots, N,
    \end{aligned}
\end{equation}

where and $p_i$ and $q_j$ are the marginal distributions of the samples from \SBone and \SBtwo respectively.
The optimal pair of points is then constructed by first selecting the samples $\mathbf{x}_1$ and then finding the corresponding $\mathbf{x}_2$ using the optimal transport map.

A point to consider is the form of the cost function $c$.
While we have found that the simply using the Euclidean distance works well, it requires transforming the features to have the same scale.
This is because the cost of transforming one feature into another is often ill-defined.
We have found that using quantile-based scalar which maps each feature to a normal distribution works well, especially when there is no prior knowledge.
It should be noted that the transformed features are only used for the cost matrix and the original features are used for the interpolation.

One drawback of using this method is that while we do not have to train any generative model, the optimal transport problem is itself computationally expensive to solve.
Current exact methods use the Hungarian algorithm~\cite{https://doi.org/10.1002/nav.3800020109}, which has a time complexity of $O(N^3)$.
Thus, it is typically infeasible to solve the optimal transport problem for the entirety of the sidebands.
We propose therefore to use a batched approach, whereby we iteratively build the template using subsets of the sidebands.
We found the variation in the generated template to be small once we used a large enough batch size ($\sim2500$), which was well within our computational resources.

The full \oli algorithm is summarized in Algorithm~\ref{alg:oli}.

\begin{algorithm}
    \caption{Pseudocode for the \oli method.}
    \label{alg:oli}
    \begin{algorithmic}
        \State \textbf{Input:} Batch size $B$, Number of batches $N$, Fitted resonant likelihood $p(m)$, Feature scaler $S$
        \State Template $ T \gets \varnothing$ \Comment{Start with empty template}
        \For{$i$ in range $N$} \Comment{Loop through the batches}
        \State $m_1, x_1 \sim$  \SBone \Comment{Sample batch from \SBone}
        \State $m_2, x_2 \sim$  \SBtwo \Comment{Sample batch from \SBtwo}
        \State $C \gets \text{compute\_cost}(S(x_1), S(x_2))$ \Comment{Compute the cost matrix using scaled features}
        \State $\Gamma^* \gets \text{solve\_optimal\_transport}(C)$ \Comment{Solve the optimal transport problem}
        \State $P \gets \text{arg\_max}(\Gamma^*)$ \Comment{Find the permutation for \SBtwo}
        \State $x_2 \gets x_2[P], m_2 \gets m_2[P]$ \Comment{Reorder \SBtwo}
        \State $m_t \sim p(m)$ \Comment{Sample batch of mass values in the SR}
        \State $x_t \gets x_1 + \left( \frac{m_t - m_1}{m_2 - m_1} (x_2 - x_1) \right)$ \Comment{Sample features using linear interpolation}
        \State $T \gets T \cup (m_t, x_t)$ \Comment{Add the batch to the template}
        \EndFor
    \end{algorithmic}
\end{algorithm}

We use the \texttt{POT} package~\cite{pot} to solve the optimal transport problem in a batched manner.
We prepare 100 batches of 5000 samples from \SBone and \SBtwo and solve the optimal transport problem for each batch.
All $(m_1, \mathbf{x}_1)$ in the batch sampled from \SBone are paired with $(m_3, \mathbf{x}_3)$ using the optimal map $\Gamma^*$ computed using that batch.
Thereafter, masses $m_2$ are sampled from the KDE and the SR template is formed using~\autoref{eq:linear_interp}.

%% file: tex/results.tex
To perform anomaly detection using \oli, we use the widely established method of CWoLa~\cite{cwola}.
Here, a classifier is trained to distinguish between the data drawn from the \SR and the template.
If the template is a good representation of the background, and the \SR is a mixture of real background and a some signal samples, the optimal classifier for this task is also the optimal classifier for distinguishing between the background and signal.

While it is not expected that \oli will outperform existing methods with more complex template building schemes, it is far more computationally efficient and robust.
Furthermore, the \oli template generation can be run on a single CPU, while efficient training of neural networks typically requires modern GPUs. 
To benchmark its performance, we compare \oli with other data-driven methods: (1) \FfF~\cite{curtainsf4f}, where normalising flows are used to construct the template, and (2) \cwola~\cite{cwola} where the SB1 and SB2 data are used directly as the template.
Note that all the template-based methods use \cwola to train a classifier between the generated (or sampled) template and data.
We label the original sideband-directly-as template approach from Ref.~\cite{cwola} as `Simple \cwola' to distinguish it from how it is used in the other methods.

While the Simple-Template method is the fastest of the three, it is expected to perform the worst as the classifier can pick up on variables which are highly correlated with the mass.
We hypothesize that we can improve on this performance using \oli, without requiring the arduous training times of \FfF.

Boosted Decision Trees (BDTs) have been shown to be very effective in anomaly detection using CWoLa~\cite{backtoroots,Freytsis:2023cjr}.
For all three methods, we use BDTs as opposed to neural networks for the classifier to further reduce the computational cost.
We used the \texttt{scikit-learn} package~\cite{scikitlearn} to grow an ensemble classifier of 50 Histogram-Gradient BDTs, each grown maximally until early stopping based on a separate 10\% hold out validation set.
For the template building methods, we used the standard four parameter di-jet fit to produce $p(m_{JJ})$~\cite{ATLAS-CONF-2023-022}.

Before assessing the quality of the templates, we compare the template generation time of 500000 events using \oli and \FfF in~\autoref{tab:time} (the simple-template generation takes no time).
We used a batch size of 5000 for \oli. \oli requires only 10 minutes to generate the template on a single CPU -- a factor of 15 times faster than \FfF, which first needs to train the two flows, then perform the morphing using GPU acceleration with each step of the process. 

\begin{table}[hbpt]
    \centering
    \caption{Comparison of the time taken for template generation using \oli and \FfF for one SR. For \FfF, this also includes the training time.}
    \label{tab:time}
    \scalebox{1.}{
        \begin{tabular}{lrr}
            \toprule
            Method & Device & Time (mins) \\
            \midrule
            \oli   & CPU    & 10 \\
            \FfF  & GPU    & 181 \\
            \bottomrule
        \end{tabular}}
\end{table}

\begin{figure}[hbpt]
\centering
\includegraphics[width=0.68\textwidth]{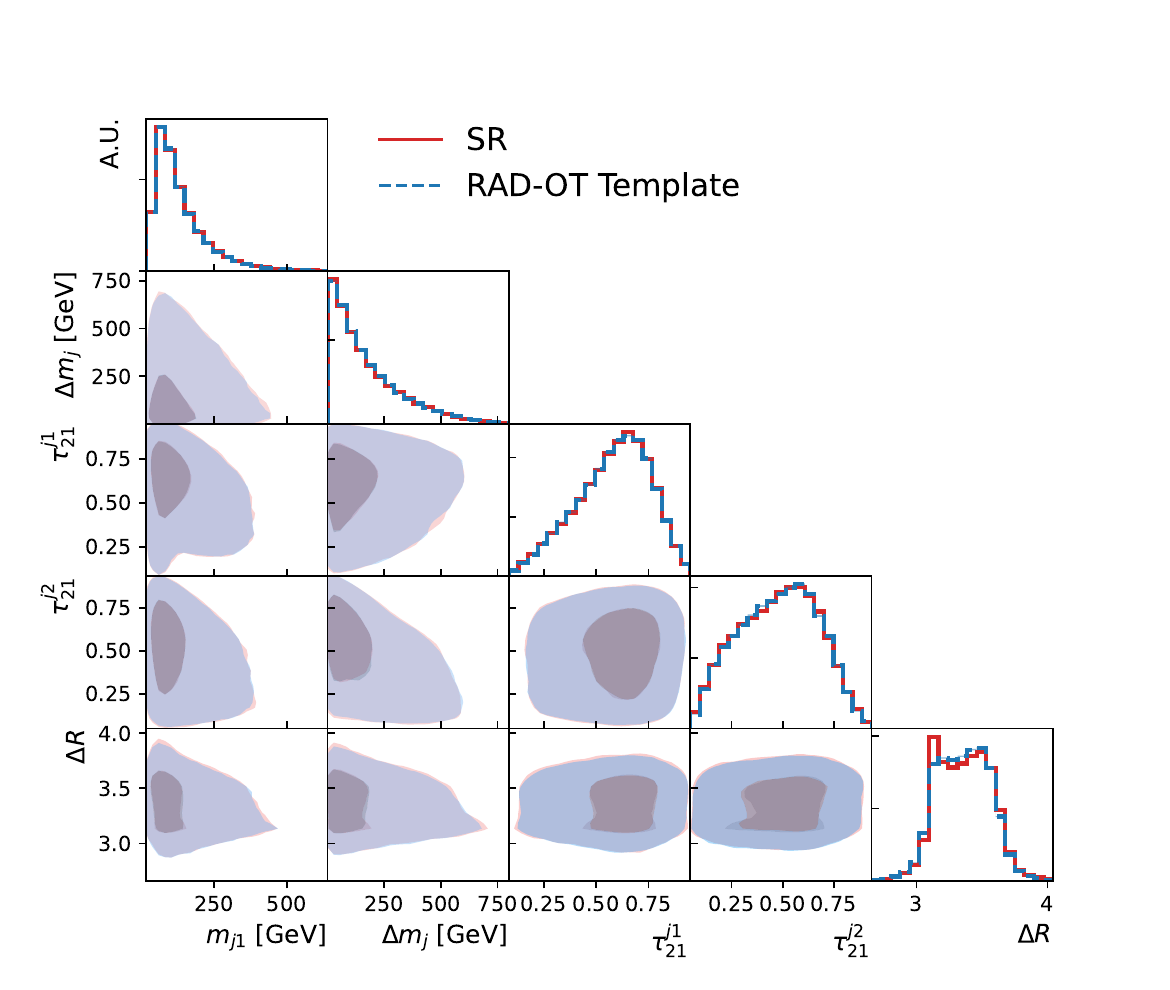}
\caption{
The template is generated using the SR $3300\leq\mjj<3700$~GeV, and  sideband regions $2900\leq\mjj<3300$~GeV and $3700\leq\mjj<4100$~GeV with no signal injected.
The diagonal elements show the marginal distributions of the features in the SR, while the off-diagonal elements show the correlations between the features.
The true data is shown in red, while the interpolated template is shown in blue.
\label{fig:template}
}
\end{figure}

\begin{figure}[hbpt]
\centering
\scalebox{0.75}{
\includegraphics[width=1.\textwidth]{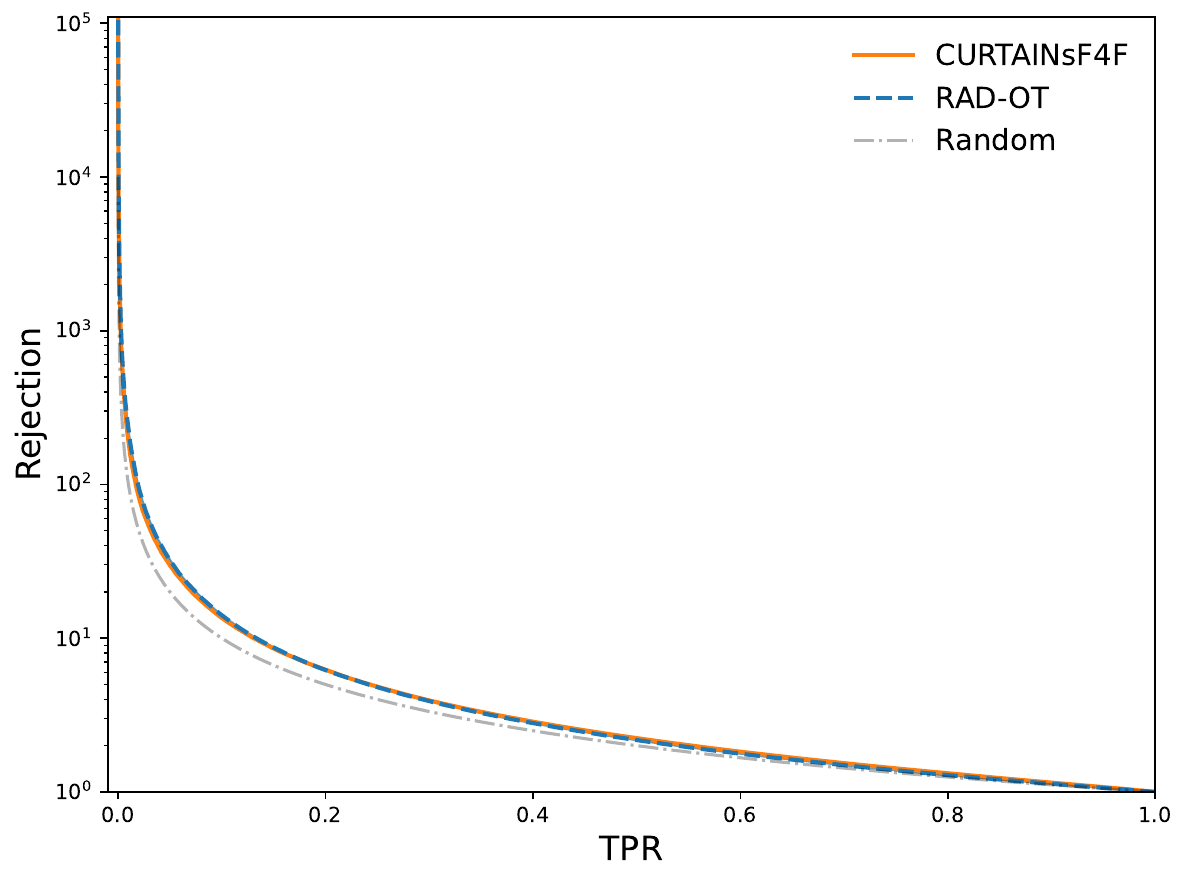}}
\caption{A Receiver Operating Characteristic (ROC) curve showing the trade-off between the true positive rate (TPR) and inverse false positive rate (1/FPR $\equiv$ rejection) for the template generated by \oli (blue) and \FfF (orange); the Random line is the case of TPR = FPR.  There is no signal here - the TPR is the probability of correctly classifying the data in the SR as such while the FPR is the probability of classifying the template data as target data.
}
\label{fig:template-roc}
\end{figure}

Next, we evaluate the template quality of \oli qualitatively by comparing the contour plots of the template and the target data in ~\autoref{fig:template}.
The marginal distributions of the features in the SR, and the correlations thereof, are captured well, with only slight mismodelling in $\Delta \text{R}$.
We quantify the \oli accuracy by training an ensemble of BDTs to distinguish between the target and the template data.
~\autoref{fig:template-roc} shows the ROC plot of this test.
An Area-Under-the-Curve (AUC) score of 0.5 indicates that the classifier cannot distinguish between these two datasets, which is our goal.
The \oli template achieves an AUC score of $0.53\pm0.01$ which is comparable to the AUC score of $0.53\pm0.01$ achieved by the \FfF method.
The values represent the mean and the standard deviation of the AUCs from 5 independent classifiers (initiated with different random seeds) trainings on the same data.

\begin{figure}[hbpt]
    \centering
    \scalebox{0.49}{
        \begin{subfigure}{1.\textwidth}
            \centering
            \includegraphics[width=\textwidth]{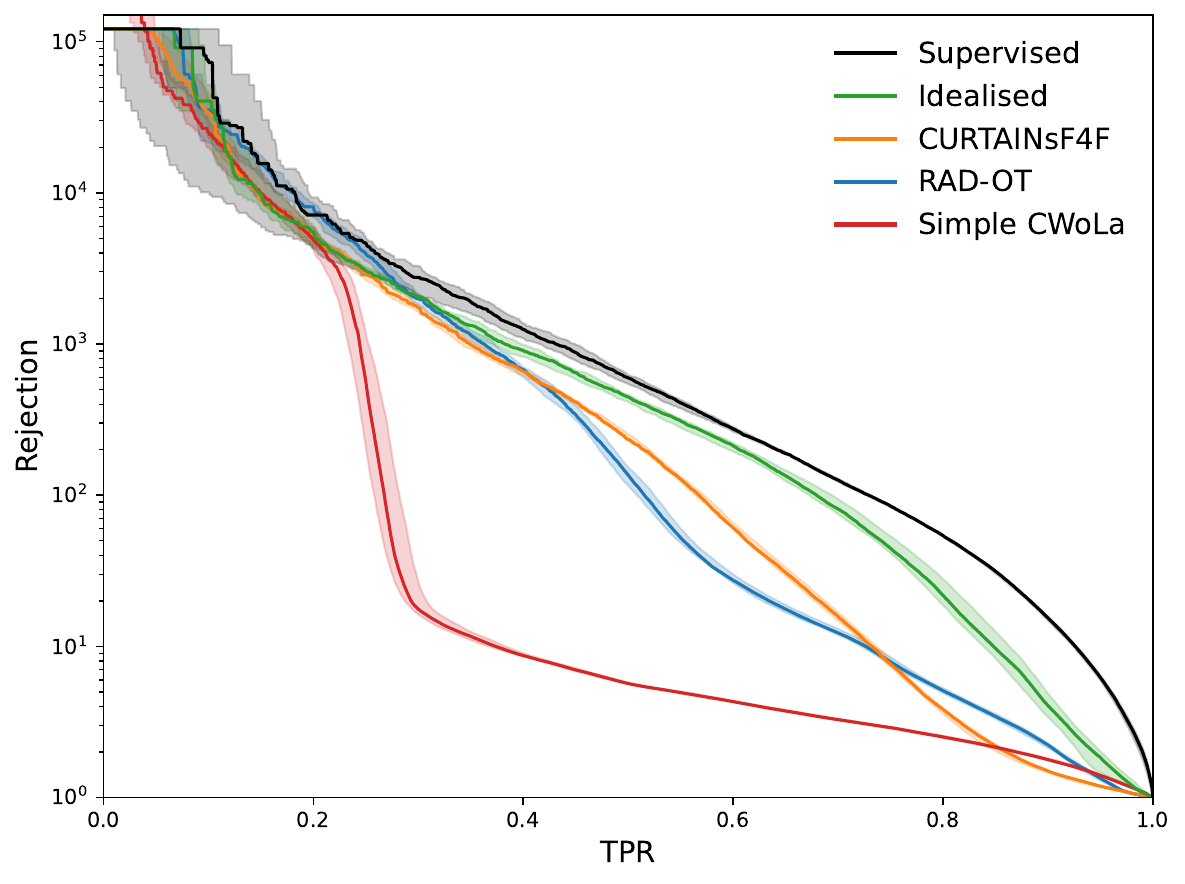}
        \end{subfigure}%
    }
    \scalebox{0.49}{
        \begin{subfigure}{1.\textwidth}
            \centering
            \includegraphics[width=\textwidth]{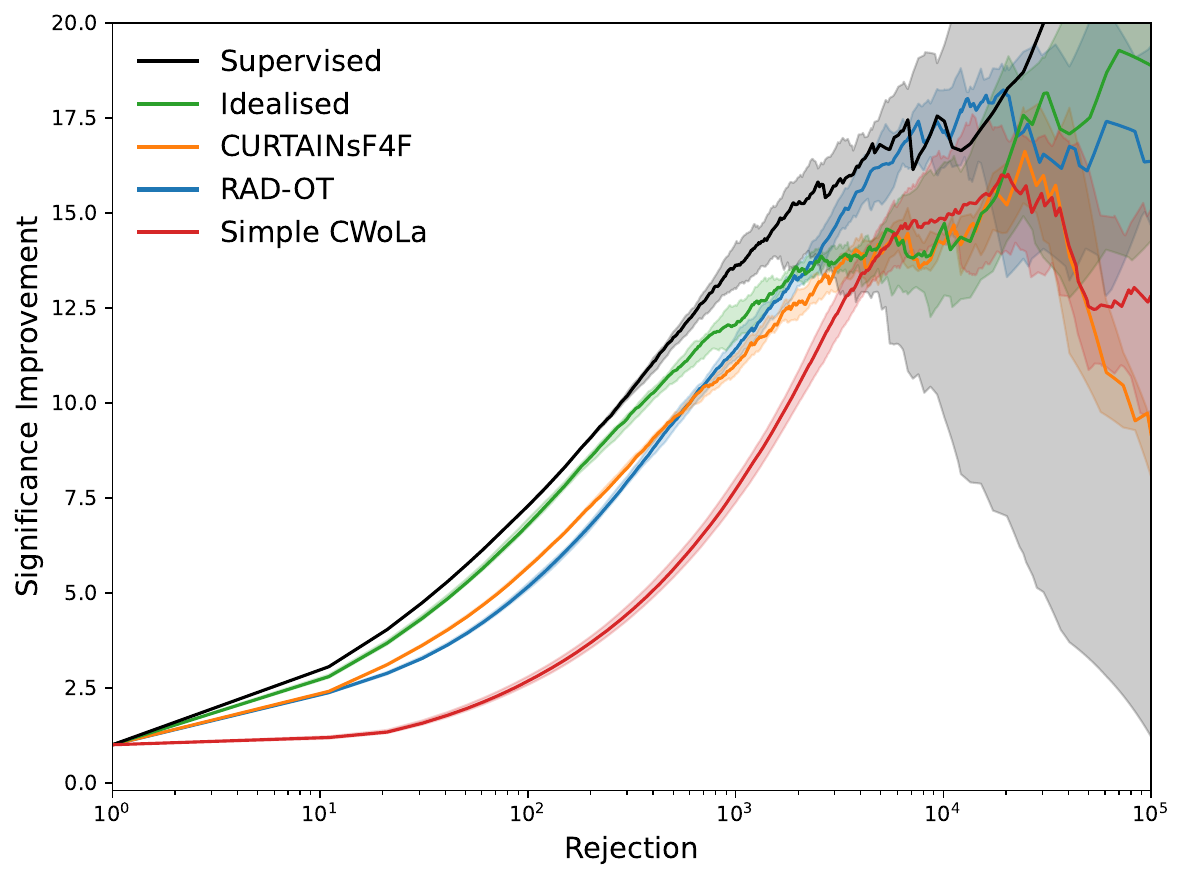}
        \end{subfigure}
    }
    \caption{Background rejection as a function of signal efficiency (left) and significance improvement as a function of background rejection (right) for \oli~(blue), \cwola~(red), Supervised~(black), Idealised~(green), and \FfF (orange).
        All classifiers are trained on the sample with 3,000 injected signal events, using a SR $3300\leq\mjj<3700$~GeV.
        The lines show the mean value of fifty classifier trainings with different random seeds, with the shaded band covering 68\% uncertainty.  The same events are used in each training; only the initialization of the machine learning varies.
        A Supervised classifier and Idealised classifiers are shown for reference.
    }
    \label{fig:sic_roc3k}
\end{figure}

After demonstrating that \oli accurately models the background, we now show how well the method can improve the significance of a signal in the SR.
For this we add 3000 signal events to the dataset ($S/\sqrt{B}=3$).
For all methods, we use 5-fold cross-validation to train the classifiers.
~\autoref{fig:sic_roc3k} shows the ROC curve and significance improvement characteristic (SIC = TPR/$\sqrt{\text{FPR}}$) versus rejection factor.
As further benchmarks, we show the performance of a supervised classifier and an idealised classifier, as in previous works.
All classifiers are made using the same ensemble of 50 BDTs but differ in the data and labels selected for training.   
The supervised classifier is trained with true signal and background labels using data from the SR.
This provides an upper bound on the achievable classification performance on the dataset.
The idealised classifier is also trained using data from the SR but we flip half of the background labels. 
This sets the limit on the performance that can be achieved with a perfect background template and noisy labels. \oli performs competitively with \FfF, achieving a SIC of $\sim 12$ at a rejection factor of 1000.
Crucially, \oli outperforms Simple \cwola, meaning that the simple method of linearly interpolating between the sidebands is an effective strategy for removing the $m_{JJ}$ dependence and enabling more sensitive anomaly detection.

\begin{figure}[hbpt]
\begin{subfigure}{0.5\textwidth}
    \centering
    \scalebox{0.99}{
        \includegraphics[width=1.\textwidth]{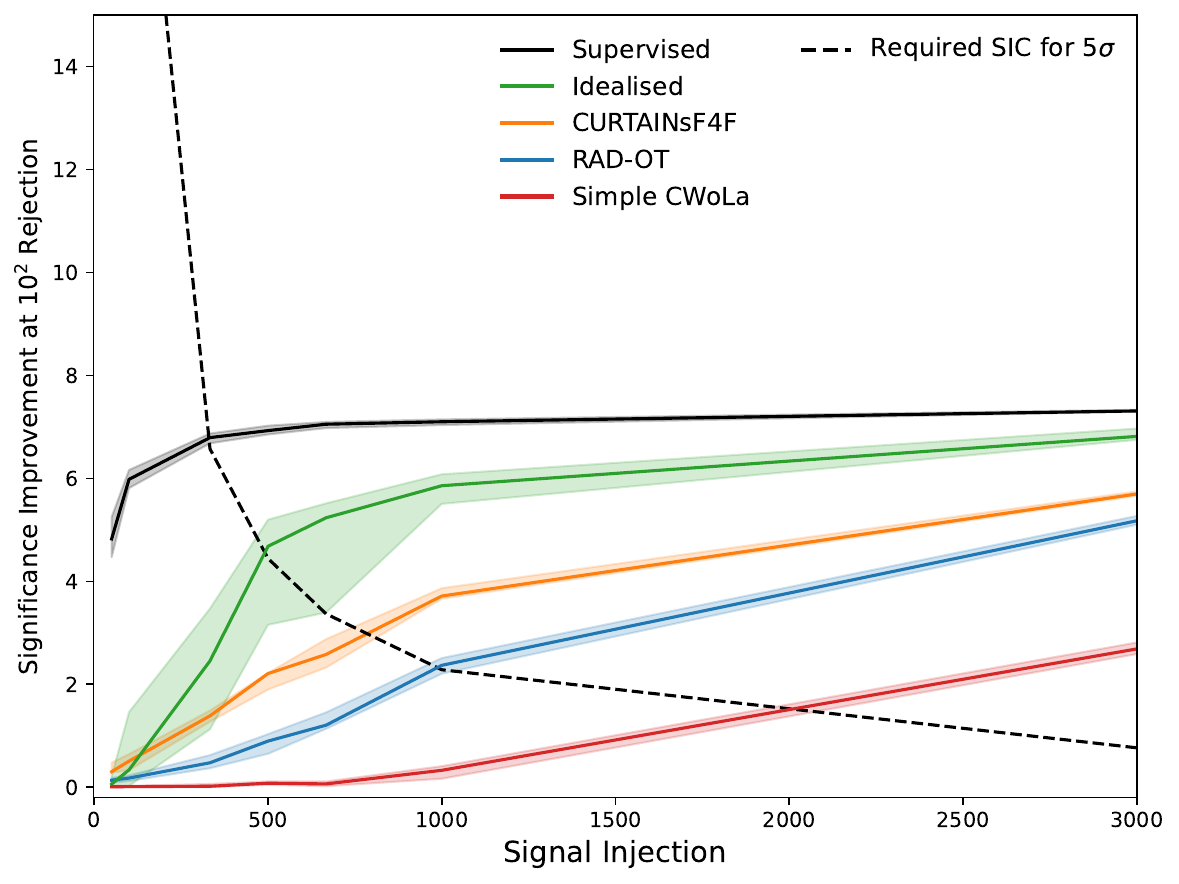}}
\end{subfigure}
\begin{subfigure}{0.5\textwidth}
    \centering
    \scalebox{0.99}{
        \includegraphics[width=1.\textwidth]{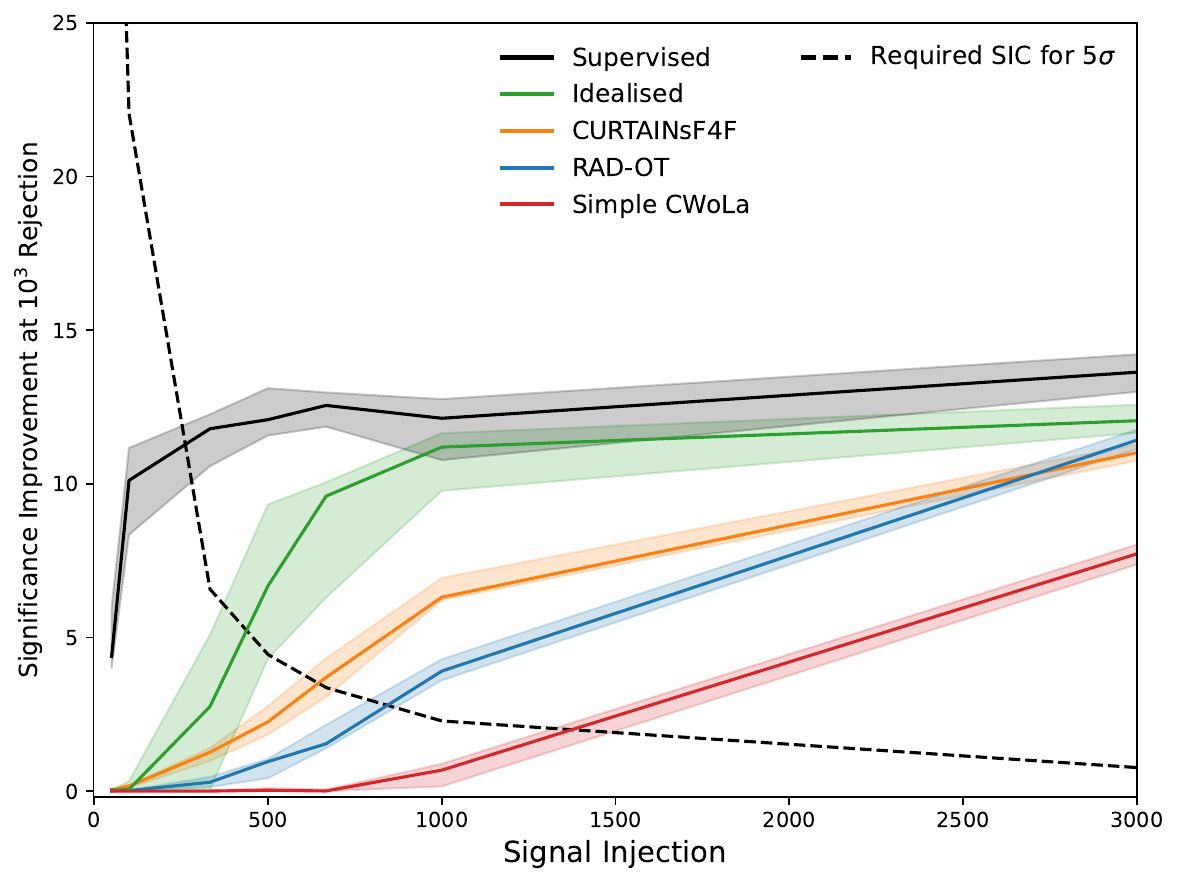}}
\end{subfigure}

    \caption{Significance improvement at a background rejection of $10^2$ (left) and $10^3$ (right) as a function of signal events in the SR $3300\leq\mjj<3700$~GeV,  for \oli~(blue), \FfF~(orange) Idealised~(green), and Supervised~(black).
        The lines show the mean value of 5 BDT trainings with different random seeds, with the shaded band covering 68\% uncertainty.
        A Supervised classifier and Idealised classifiers are shown for reference.
        The required SIC across the doping levels for a discovery is shown as a dashed line.
    }
    \label{fig:sic_vs_sig}
\end{figure}

We also track the sensitivity of the method to different levels of signal injection and calculate the SIC at a rejection factor of 1000 which is shown in ~\autoref{fig:sic_vs_sig}.
The required SIC for a \emph{discovery} is shown as a dashed line.
Here, we use $\frac{S}{\sqrt{B}} = 5$ as a discovery threshold.
We see that \oli performs better than standalone \cwola and is able to \emph{discover} a signal with as few as $\lesssim 700$ signal events in the SR, which corresponds to an initial $\frac{S}{\sqrt{B}} \sim 2$.

\begin{figure}[hbpt]
    \centering
    \includegraphics[width=0.49\textwidth]{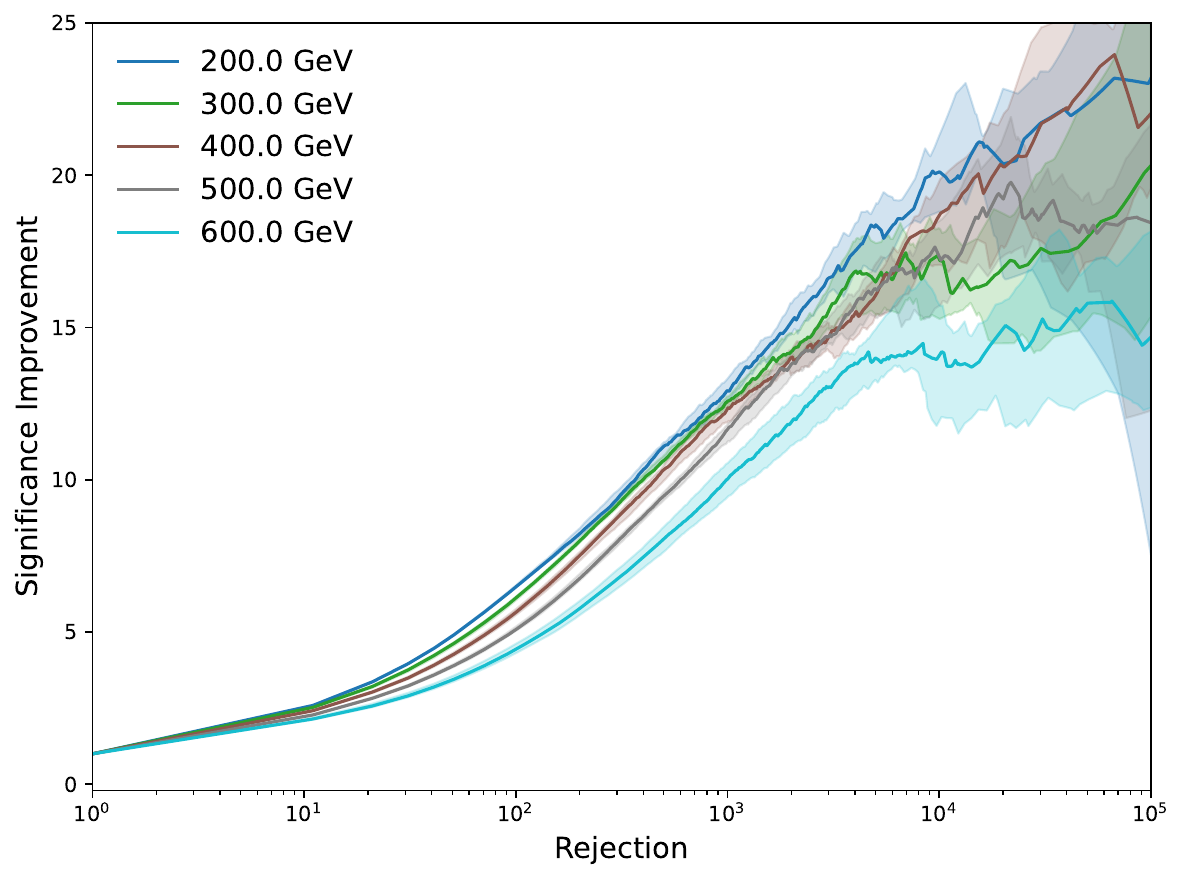}
    \includegraphics[width=0.49\textwidth]{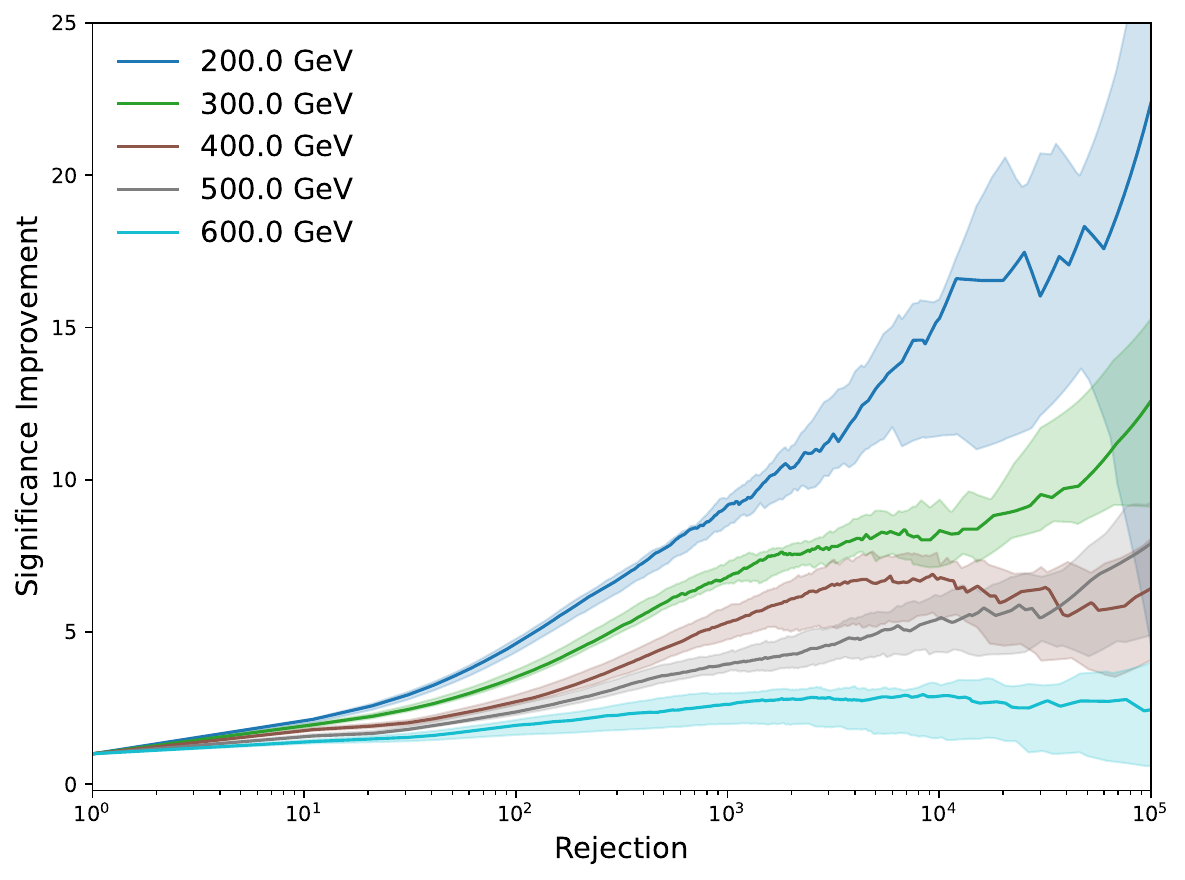}
    \caption{The SIC vs rejection factor for different SR widths using \oli on datasets with 3000 (left) and 1000 (right) injected signal samples. All SRs are centred on 3500~GeV. 
    }
    \label{fig:datarocsicwidth}
\end{figure}

The main assumptions of the \oli method is that the conditional probability density of the classifier features varies linearly along the optimal transport path connecting the resonant feature.
While this assumption may hold in small regions of the phase (i.e. narrow signal regions and sidebands), it is crucial to investigate the effect of the window widths on the performance of the method.
To do this, we fix the sideband width and vary the SR window width to see how the performance of the method changes.
~\autoref{fig:datarocsicwidth} shows and SIC vs rejection factor for different SR window widths for the \oli method.
As expected, the performance of \oli generally decreases with increasing SR width, as the linear approximation between the features and the resonant feature no longer holds true, leading to a worse template, and hence a worse classifier performance.
With 3000 injected signal events, even with a \SR with of 600~GeV, \oli is still able to reach a SIC of $\sim 10$ at a rejection factor of 1000.
However, with only 1000 injected signal events, there is a notable performance drop with even 300 GeV, highlighting the sensitivity of this method to wider signal regions. 

%% file: tex/discussion.tex
In this work, we develop a method for generating templates in resonant anomaly detection using optimal transport and linear interpolation to enhance stability and reduce generation time compared to previous methods.
In order to ensure that the interpolation paths are meaningful, we match pairs of samples from each sideband using a mini-batched optimal transport solution.  This approach assumes that the conditional probability density of the classifier features varies linearly along the optimal transport path connecting the resonant feature.  While this assumption may not be exact, it is a reasonable first order approximation to apply to narrow signal regions.
We validated this approach on the LHCO dataset and showed competitive performance with more complex template generation methods, based on neural networks, that take an order of magnitude longer to train.  Our new \oli method provides a complementary approach to existing methods and may enable faster sweeps of signal regions when computational challenges are limiting.  It would also be interesting to explore how the precision of \oli scales with the number of features and the amount of available data, where it may provide an advantage with limited data as no (neural network) training is required.

%% file: includes/acknowledgement.tex
DS, and TG acknowledge funding through the SNSF Sinergia grant CRSII5\_193716 ``Robust Deep Density Models for High-Energy Particle Physics and Solar Flare Analysis (RODEM)''
and the SNSF project grant 200020\_212127 ``At the two upgrade frontiers: machine learning and the ITk Pixel detector''.
ML would like to acknowledge individual funding acquired through the Swiss Government Excellence Scholarships for Foreign Scholars.
BN is supported by the U.S. Department of Energy (DOE), Office of Science under contract DE-AC02-05CH11231.

%% file: main.bib
@article{drapes,
author = {Sengupta, Debajyoti and Leigh, Matthew and Raine, John and Klein, Samuel and Golling, Tobias},
year = {2024},
month = {04},
pages = {},
title = {Improving new physics searches with diffusion models for event observables and jet constituents},
volume = {2024},
journal = {Journal of High Energy Physics},
doi = {10.1007/JHEP04(2024)109}
}

@article{curtainsf4f,
    author = "Sengupta, Debajyoti and Klein, Samuel and Raine, John Andrew and Golling, Tobias",
    title = "{CURTAINs Flows For Flows: Constructing Unobserved Regions with Maximum Likelihood Estimation}",
    eprint = "2305.04646",
    archivePrefix = "arXiv",
    primaryClass = "hep-ph",
    month = "5",
    year = "2023"
}

@techreport{CMS-PAS-EXO-22-026,
      collaboration = "CMS",
      title         = "{Model-agnostic search for dijet resonances with anomalous
                       jet substructure in proton-proton collisions at $\sqrt{s}$
                       = 13 TeV}",
      institution   = "CERN",
      reportNumber  = "CMS-PAS-EXO-22-026",
      address       = "Geneva",
      year          = "2024",
      url           = "https://cds.cern.ch/record/2892677",
}

@techreport{ATLAS-CONF-2023-022,
      collaboration = "ATLAS",
      title         = "{Search for new phenomena in two-body invariant mass
                       distributions using unsupervised machine learning for
                       anomaly detection at $\sqrt{s} = 13$ TeV with the ATLAS
                       detector}",
      institution   = "CERN",
      reportNumber  = "ATLAS-CONF-2023-022",
      address       = "Geneva",
      year          = "2023",
      url           = "https://cds.cern.ch/record/2859329",
      
}

@article{ATLAS:2023azi,
    author = "{ATLAS Collaboration}",
    collaboration = "ATLAS",
    title = "{Anomaly detection search for new resonances decaying into a Higgs boson and a generic new particle $X$ in hadronic final states using $\sqrt{s} = 13$ TeV $pp$ collisions with the ATLAS detector}",
    eprint = "2306.03637",
    archivePrefix = "arXiv",
    primaryClass = "hep-ex",
    reportNumber = "CERN-EP-2023-045",
    month = "6",
    year = "2023"
}

@article{Pettee:2023zra,
    author = "Pettee, Mariel and Thanvantri, Sowmya and Nachman, Benjamin and Shih, David and Buckley, Matthew R. and Collins, Jack H.",
    title = "{Weakly-Supervised Anomaly Detection in the Milky Way}",
    eprint = "2305.03761",
    archivePrefix = "arXiv",
    primaryClass = "astro-ph.GA",
    month = "5",
    year = "2023"
}

@article{Shih:2023jfv,
    author = "Shih, David and Buckley, Matthew R. and Necib, Lina",
    title = "{Via Machinae 2.0: Full-Sky, Model-Agnostic Search for Stellar Streams in Gaia DR2}",
    eprint = "2303.01529",
    archivePrefix = "arXiv",
    primaryClass = "astro-ph.GA",
    month = "3",
    year = "2023"
}

@article{ATLAS:2020iwa,
    author = "{ATLAS Collaboration}",
    collaboration = "ATLAS",
    title = "{Dijet resonance search with weak supervision using $\sqrt{s}=13$ TeV $pp$ collisions in the ATLAS detector}",
    eprint = "2005.02983",
    archivePrefix = "arXiv",
    primaryClass = "hep-ex",
    reportNumber = "CERN-EP-2020-062",
    doi = "10.1103/PhysRevLett.125.131801",
    journal = "Phys. Rev. Lett.",
    volume = "125",
    number = "13",
    pages = "131801",
    year = "2020"
}

@article{Amram_2021,
	doi = {10.1007/jhep01(2021)153},
	url = {https://doi.org/10.1007%2Fjhep01%282021%29153},
	year = 2021,
	month = jan,
	publisher = {Springer Science and Business Media {LLC}},
	volume = {2021},
	number = {1},
	author = {Oz Amram and Cristina Mantilla Suarez},
	title = {Tag N' Train: a technique to train improved classifiers on unlabeled data},
	journal = {Journal of High Energy Physics}
}

@article{Sengupta:2024ezl,
    author = "Sengupta, Debajyoti and Mulligan, Stephen and Shih, David and Raine, John Andrew and Golling, Tobias",
    title = "{SkyCURTAINs: Model agnostic search for Stellar Streams with Gaia data}",
    eprint = "2405.12131",
    archivePrefix = "arXiv",
    primaryClass = "astro-ph.GA",
    month = "5",
    year = "2024"
}

@article{feta,
    author = "Golling, Tobias and Klein, Samuel and Mastandrea, Radha and Nachman, Benjamin",
    title = "{Flow-enhanced transportation for anomaly detection}",
    eprint = "2212.11285",
    archivePrefix = "arXiv",
    primaryClass = "hep-ph",
    doi = "10.1103/PhysRevD.107.096025",
    journal = "Phys. Rev. D",
    volume = "107",
    number = "9",
    pages = "096025",
    year = "2023"
}

@article{Metodiev:2023izu,
    author = "Metodiev, Eric M. and Thaler, Jesse and Wynne, Raymond",
    title = "{Anomaly Detection in Collider Physics via Factorized Observables}",
    eprint = "2312.00119",
    archivePrefix = "arXiv",
    primaryClass = "hep-ph",
    reportNumber = "MIT-CTP/5644",
    month = "11",
    year = "2023"
}

@article{Golling:2023yjq,
    author = "Golling, Tobias and Kasieczka, Gregor and Krause, Claudius and Mastandrea, Radha and Nachman, Benjamin and Raine, John Andrew and Sengupta, Debajyoti and Shih, David and Sommerhalder, Manuel",
    title = "{The interplay of machine learning-based resonant anomaly detection methods}",
    eprint = "2307.11157",
    archivePrefix = "arXiv",
    primaryClass = "hep-ph",
    doi = "10.1140/epjc/s10052-024-12607-x",
    journal = "Eur. Phys. J. C",
    volume = "84",
    number = "3",
    pages = "241",
    year = "2024"
}

@article{Buhmann:2023acn,
    author = "Buhmann, Erik and Ewen, Cedric and Kasieczka, Gregor and Mikuni, Vinicius and Nachman, Benjamin and Shih, David",
    title = "{Full phase space resonant anomaly detection}",
    eprint = "2310.06897",
    archivePrefix = "arXiv",
    primaryClass = "hep-ph",
    doi = "10.1103/PhysRevD.109.055015",
    journal = "Phys. Rev. D",
    volume = "109",
    number = "5",
    pages = "055015",
    year = "2024"
}

@article{Freytsis:2023cjr,
    author = "Freytsis, Marat and Perelstein, Maxim and San, Yik Chuen",
    title = "{Anomaly detection in the presence of irrelevant features}",
    eprint = "2310.13057",
    archivePrefix = "arXiv",
    primaryClass = "hep-ph",
    doi = "10.1007/JHEP02(2024)220",
    journal = "JHEP",
    volume = "02",
    pages = "220",
    year = "2024"
}

@article{https://doi.org/10.1002/nav.3800020109,
author = {Kuhn, H. W.},
title = {The Hungarian method for the assignment problem},
journal = {Naval Research Logistics Quarterly},
volume = {2},
number = {1-2},
pages = {83-97},
eprint = {https://onlinelibrary.wiley.com/doi/pdf/10.1002/nav.3800020109},
year = {1955},
}

@article{Fraser:2021lxm,
    author = "Fraser, Katherine and Homiller, Samuel and Mishra, Rashmish K. and Ostdiek, Bryan and Schwartz, Matthew D.",
    title = "{Challenges for unsupervised anomaly detection in particle physics}",
    eprint = "2110.06948",
    archivePrefix = "arXiv",
    primaryClass = "hep-ph",
    doi = "10.1007/JHEP03(2022)066",
    journal = "JHEP",
    volume = "03",
    pages = "066",
    year = "2022"
}

@article{CrispimRomao:2020ejk,
    author = "Crispim Rom\~ao, M. and Castro, N. F. and Milhano, J. G. and Pedro, R. and Vale, T.",
    title = "{Use of a generalized energy Mover\textquoteright{}s distance in the search for rare phenomena at colliders}",
    eprint = "2004.09360",
    archivePrefix = "arXiv",
    primaryClass = "hep-ph",
    doi = "10.1140/epjc/s10052-021-08891-6",
    journal = "Eur. Phys. J. C",
    volume = "81",
    number = "2",
    pages = "192",
    year = "2021"
}

@article{Park:2022zov,
    author = "Park, Sang Eon and Harris, Philip and Ostdiek, Bryan",
    title = "{Neural embedding: learning the embedding of the manifold of physics data}",
    eprint = "2208.05484",
    archivePrefix = "arXiv",
    primaryClass = "hep-ph",
    doi = "10.1007/JHEP07(2023)108",
    journal = "JHEP",
    volume = "07",
    pages = "108",
    year = "2023"
}

@article{Craig:2024rlv,
    author = "Craig, Nathaniel and Howard, Jessica N. and Li, Hancheng",
    title = "{Exploring Optimal Transport for Event-Level Anomaly Detection at the Large Hadron Collider}",
    eprint = "2401.15542",
    archivePrefix = "arXiv",
    primaryClass = "hep-ph",
    month = "1",
    year = "2024"
}

@article{curtains,
    author = "Raine, John Andrew and Klein, Samuel and Sengupta, Debajyoti and Golling, Tobias",
    title = "{CURTAINs for your sliding window: Constructing unobserved regions by transforming adjacent intervals}",
    eprint = "2203.09470",
    archivePrefix = "arXiv",
    primaryClass = "hep-ph",
    doi = "10.3389/fdata.2023.899345",
    journal = "Front. Big Data",
    volume = "6",
    pages = "899345",
    year = "2023"
}

@article{ATLAS:2008xda,
    author = "{ATLAS Collaboration}",
    collaboration = "ATLAS",
    title = "{The ATLAS Experiment at the CERN Large Hadron Collider}",
    doi = "10.1088/1748-0221/3/08/S08003",
    journal = "JINST",
    volume = "3",
    pages = "S08003",
    year = "2008"
}

@article{CMS:2008xjf,
    author = "{CMS Collaboration}",
    collaboration = "CMS",
    title = "{The CMS Experiment at the CERN LHC}",
    doi = "10.1088/1748-0221/3/08/S08004",
    journal = "JINST",
    volume = "3",
    pages = "S08004",
    year = "2008"
}

@article{Aarrestad:2021oeb,
    author = "Aarrestad, Thea and others",
    title = "{The Dark Machines Anomaly Score Challenge: Benchmark Data and Model Independent Event Classification for the Large Hadron Collider}",
    eprint = "{2105.14027}",
    archivePrefix = "arXiv",
    primaryClass = "hep-ph",
    reportNumber = "FERMILAB-PUB-21-285-CMS",
    doi = "10.21468/SciPostPhys.12.1.043",
    journal = "SciPost Phys.",
    volume = "12",
    number = "1",
    pages = "043",
    year = "2022"
}

@dataset{LHCOlympics,
   author       = "Gregor Kasieczka and Benjamin Nachman and David Shih",
   year         = "2019",
   title       = "{Official Datasets for LHC Olympics 2020 Anomaly Detection Challenge (Version v6)}",
   publisher    = {Zenodo},
   doi = "10.5281/zenodo.4536624",
}

@article{Kasieczka:2021xcg,
    author = "Kasieczka, Gregor and others",
    title = "{The LHC Olympics 2020 a community challenge for anomaly detection in high energy physics}",
    eprint = "2101.08320",
    archivePrefix = "arXiv",
    primaryClass = "hep-ph",
    doi = "10.1088/1361-6633/ac36b9",
    journal = "Rept. Prog. Phys.",
    volume = "84",
    number = "12",
    pages = "124201",
    year = "2021"
}

@article{Sjostrand:2014zea,
    author = {Sj\"ostrand, Torbj\"orn and Ask, Stefan and Christiansen, Jesper R. and Corke, Richard and Desai, Nishita and Ilten, Philip and Mrenna, Stephen and Prestel, Stefan and Rasmussen, Christine O. and Skands, Peter Z.},
    title = "{An introduction to PYTHIA 8.2}",
    eprint = "1410.3012",
    archivePrefix = "arXiv",
    primaryClass = "hep-ph",
    reportNumber = "LU-TP-14-36, MCNET-14-22, CERN-PH-TH-2014-190, FERMILAB-PUB-14-316-CD, DESY-14-178, SLAC-PUB-16122",
    doi = "10.1016/j.cpc.2015.01.024",
    journal = "Comput. Phys. Commun.",
    volume = "191",
    pages = "159--177",
    year = "2015"
}

@article{deFavereau:2013fsa,
    author = "de Favereau, J. and Delaere, C. and Demin, P. and Giammanco, A. and Lema\^\i{}tre, V. and Mertens, A. and Selvaggi, M.",
    title = "{DELPHES 3, A modular framework for fast simulation of a generic collider experiment}",
    eprint = "1307.6346",
    archivePrefix = "arXiv",
    primaryClass = "hep-ex",
    doi = "10.1007/JHEP02(2014)057",
    journal = "JHEP",
    volume = "02",
    pages = "057",
    year = "2014"
}

@article{AntiKt,
  author       = {Cacciari, Matteo and Salam, Gavin P and Soyez, Gregory},
  year         = {2008},
  doi          = {10.1088/1126-6708/2008/04/063},
  journal      = {JHEP},
  pages        = {063},
  title        = {The anti-kt jet clustering algorithm},
  volume       = {04},
}

@article{Cacciari:2011ma,
    author = "Cacciari, Matteo and Salam, Gavin P. and Soyez, Gregory",
    title = "{FastJet User Manual}",
    eprint = "{1111.6097}",
    archivePrefix = "arXiv",
    primaryClass = "hep-ph",
    reportNumber = "CERN-PH-TH-2011-297",
    doi = "10.1140/epjc/s10052-012-1896-2",
    journal = "Eur. Phys. J. C",
    volume = "72",
    pages = "1896",
    year = "2012"
}

@article{nsubjettiness,
   title={Identifying boosted objects with N-subjettiness},
   volume={2011},
   ISSN={1029-8479},
   url={http://dx.doi.org/10.1007/JHEP03(2011)015},
   DOI={10.1007/jhep03(2011)015},
   number={3},
   journal={Journal of High Energy Physics},
   publisher={Springer Science and Business Media LLC},
   author={Thaler, Jesse and Van Tilburg, Ken},
   year={2011},
   month=Mar,
}

@article{anode,
    author = "Nachman, Benjamin and Shih, David",
    title = "{Anomaly Detection with Density Estimation}",
    eprint = "2001.04990",
    archivePrefix = "arXiv",
    primaryClass = "hep-ph",
    doi = "10.1103/PhysRevD.101.075042",
    journal = "Phys. Rev. D",
    volume = "101",
    pages = "075042",
    year = "2020"
}

@article{cathode,
    author = "Hallin, Anna and Isaacson, Joshua and Kasieczka, Gregor and Krause, Claudius and Nachman, Benjamin and Quadfasel, Tobias and Schlaffer, Matthias and Shih, David and Sommerhalder, Manuel",
    title = "{Classifying anomalies through outer density estimation}",
    reportNumber = "EFI-20-5, FERMILAB-PUB-21-389-T",
    doi = "10.1103/PhysRevD.106.055006",
    journal = "Phys. Rev. D",
    volume = "106",
    number = "5",
    pages = "055006",
    year = "2022"
}

@article{cwola,
    author = "Metodiev, Eric M. and Nachman, Benjamin and Thaler, Jesse",
    title = "{Classification without labels: Learning from mixed samples in high energy physics}",
    eprint = "1708.02949",
    archivePrefix = "arXiv",
    primaryClass = "hep-ph",
    reportNumber = "MIT--CTP-4922",
    doi = "10.1007/JHEP10(2017)174",
    journal = "JHEP",
    volume = "10",
    pages = "174",
    year = "2017"
}

@article{cwolabump,
    author = "Collins, Jack H. and Howe, Kiel and Nachman, Benjamin",
    title = "{Extending the search for new resonances with machine learning}",
    eprint = "1902.02634",
    archivePrefix = "arXiv",
    primaryClass = "hep-ph",
    reportNumber = "FERMILAB-PUB-18-733-T",
    doi = "10.1103/PhysRevD.99.014038",
    journal = "Phys. Rev. D",
    volume = "99",
    number = "1",
    pages = "014038",
    year = "2019"
}

@article{Stein:2020rou,
    author = "Stein, George and Seljak, Uros and Dai, Biwei",
    title = "{Unsupervised in-distribution anomaly detection of new physics through conditional density estimation}",
    booktitle = "{34th Conference on Neural Information Processing Systems}",
    eprint = "2012.11638",
    archivePrefix = "arXiv",
    primaryClass = "cs.LG",
    month = "12",
    year = "2020"
}

@article{Chen:2022suv,
    author = "Chen, Mayee F. and Nachman, Benjamin and Sala, Frederic",
    title = "{Resonant Anomaly Detection with Multiple Reference Datasets}",
    eprint = "2212.10579",
    archivePrefix = "arXiv",
    primaryClass = "hep-ph",
    month = "12",
    year = "2022"
}

@article{backtoroots,
    author = {Finke, Thorben and Hein, Marie and Kasieczka, Gregor and Kr\"amer, Michael and M\"uck, Alexander and Prangchaikul, Parada and Quadfasel, Tobias and Shih, David and Sommerhalder, Manuel},
    title = "{Back To The Roots: Tree-Based Algorithms for Weakly Supervised Anomaly Detection}",
    eprint = "2309.13111",
    archivePrefix = "arXiv",
    primaryClass = "hep-ph",
    reportNumber = "TTK-23-26",
    month = "9",
    year = "2023"
}

@article{Shih:2021kbt,
    author = "Shih, David and Buckley, Matthew R. and Necib, Lina and Tamanas, John",
    title = "{via machinae: Searching for stellar streams using unsupervised machine learning}",
    eprint = "2104.12789",
    archivePrefix = "arXiv",
    primaryClass = "astro-ph.GA",
    doi = "10.1093/mnras/stab3372",
    journal = "Mon. Not. Roy. Astron. Soc.",
    volume = "509",
    number = "4",
    pages = "5992--6007",
    year = "2021"
}

@article{Benkendorfer:2020gek,
    author = "Benkendorfer, Kees and Pottier, Luc Le and Nachman, Benjamin",
    title = "{Simulation-assisted decorrelation for resonant anomaly detection}",
    eprint = "2009.02205",
    archivePrefix = "arXiv",
    primaryClass = "hep-ph",
    doi = "10.1103/PhysRevD.104.035003",
    journal = "Phys. Rev. D",
    volume = "104",
    number = "3",
    pages = "035003",
    year = "2021"
}

@article{Collins:2018epr,
    author = "Collins, Jack H. and Howe, Kiel and Nachman, Benjamin",
    title = "{Anomaly Detection for Resonant New Physics with Machine Learning}",
    eprint = "1805.02664",
    archivePrefix = "arXiv",
    primaryClass = "hep-ph",
    reportNumber = "FERMILAB-PUB-18-180-T",
    doi = "10.1103/PhysRevLett.121.241803",
    journal = "Phys. Rev. Lett.",
    volume = "121",
    number = "24",
    pages = "241803",
    year = "2018"
}

@article{Andreassen:2020nkr,
    author = "Andreassen, Anders and Nachman, Benjamin and Shih, David",
    title = "{Simulation Assisted Likelihood-free Anomaly Detection}",
    eprint = "2001.05001",
    archivePrefix = "arXiv",
    primaryClass = "hep-ph",
    doi = "10.1103/PhysRevD.101.095004",
    journal = "Phys. Rev. D",
    volume = "101",
    number = "9",
    pages = "095004",
    year = "2020"
}

@article{Karagiorgi:2021ngt,
    author = "Karagiorgi, Georgia and Kasieczka, Gregor and Kravitz, Scott and Nachman, Benjamin and Shih, David",
    title = "{Machine Learning in the Search for New Fundamental Physics}",
    eprint = "2112.03769",
    archivePrefix = "arXiv",
    primaryClass = "hep-ph",
    month = "12",
    year = "2021"
}

@article{Kamenik:2022qxs,
    author = "Kamenik, Jernej F. and Szewc, Manuel",
    title = "{Null hypothesis test for anomaly detection}",
    eprint = "2210.02226",
    archivePrefix = "arXiv",
    primaryClass = "hep-ph",
    doi = "10.1016/j.physletb.2023.137836",
    journal = "Phys. Lett. B",
    volume = "840",
    pages = "137836",
    year = "2023"
}

@article{pot,
  author  = {RÃ©mi Flamary and Nicolas Courty and Alexandre Gramfort and Mokhtar Z. Alaya and AurÃ©lie Boisbunon and Stanislas Chambon and Laetitia Chapel and Adrien Corenflos and Kilian Fatras and Nemo Fournier and LÃ©o Gautheron and Nathalie T.H. Gayraud and Hicham Janati and Alain Rakotomamonjy and Ievgen Redko and Antoine Rolet and Antony Schutz and Vivien Seguy and Danica J. Sutherland and Romain Tavenard and Alexander Tong and Titouan Vayer},
  title   = {POT: Python Optimal Transport},
  journal = {Journal of Machine Learning Research},
  year    = {2021},
  volume  = {22},
  number  = {78},
  pages   = {1--8},
  url     = {http://jmlr.org/papers/v22/20-451.html}
}

@misc{ranode,
      title={Residual ANODE}, 
      author={Ranit Das and Gregor Kasieczka and David Shih},
      year={2023},
      eprint={2312.11629},
      archivePrefix={arXiv}}

@misc{scikitlearn,
      title={API design for machine learning software: experiences from the scikit-learn project}, 
      author={Lars Buitinck and Gilles Louppe and Mathieu Blondel and Fabian Pedregosa and Andreas Mueller and Olivier Grisel and Vlad Niculae and Peter Prettenhofer and Alexandre Gramfort and Jaques Grobler and Robert Layton and Jake Vanderplas and Arnaud Joly and Brian Holt and Gaël Varoquaux},
      year={2013},
      eprint={1309.0238},
      archivePrefix={arXiv}
}
